\documentclass[superscriptaddress, preprint, floatfix]{revtex4-1}

\usepackage{here}
\usepackage{graphicx}
\usepackage{wrapfig}
\usepackage{sidecap}

\usepackage{multirow}

\usepackage{amsmath}
\usepackage{amssymb}
\usepackage{braket}

\usepackage{hyperref}

\begin{document}
\title{Dynamics of the photoinduced insulator-to-metal transition in a nickelate film}
\date{\today}

\author{Vincent Esposito}
\affiliation{Photon Science Division, Paul Scherrer Institut, 5232 Villigen PSI, Switzerland}

\author{Laurenz Rettig}
\thanks{Current address: Fritz Haber Institute of the Max Planck Society, 14195 Berlin, Germany}
\affiliation{Photon Science Division, Paul Scherrer Institut, 5232 Villigen PSI, Switzerland}

\author{Elisabeth M. Bothschafter}
\affiliation{Photon Science Division, Paul Scherrer Institut, 5232 Villigen PSI, Switzerland}

\author{Yunpei Deng}
\affiliation{Photon Science Division, Paul Scherrer Institut, 5232 Villigen PSI, Switzerland}

\author{Christian Dornes}
\affiliation{Institute for Quantum Electronics, ETH Z\"{u}rich, 8093 Z\"{u}rich, Switzerland}

\author{Lucas Huber}
\affiliation{Institute for Quantum Electronics, ETH Z\"{u}rich, 8093 Z\"{u}rich, Switzerland}

\author{Tim Huber}
\affiliation{Institute for Quantum Electronics, ETH Z\"{u}rich, 8093 Z\"{u}rich, Switzerland}

\author{Gerhard Ingold}
\affiliation{Photon Science Division, Paul Scherrer Institut, 5232 Villigen PSI, Switzerland}

\author{Yuichi Inubushi}
\affiliation{RIKEN SPring-8 Center, Hyogo 679-5148, Japan}
\affiliation{Japan Synchrotron Radiation Research Institute, Hyogo 679-5198, Japan}

\author{Tetsuo Katayama}
\affiliation{RIKEN SPring-8 Center, Hyogo 679-5148, Japan}
\affiliation{Japan Synchrotron Radiation Research Institute, Hyogo 679-5198, Japan}

\author{Tomoya Kawaguchi}
\affiliation{Department of Materials Science and Engineering, Kyoto University, Kyoto 606-8501, Japan}

\author{Henrik Lemke}
\affiliation{Photon Science Division, Paul Scherrer Institut, 5232 Villigen PSI, Switzerland}

\author{Kanade Ogawa}
\affiliation{RIKEN SPring-8 Center, Hyogo 679-5148, Japan}
\affiliation{Japan Synchrotron Radiation Research Institute, Hyogo 679-5198, Japan}

\author{Shigeki Owada}
\affiliation{RIKEN SPring-8 Center, Hyogo 679-5148, Japan}
\affiliation{Japan Synchrotron Radiation Research Institute, Hyogo 679-5198, Japan}

\author{Milan Radovic}
\affiliation{Photon Science Division, Paul Scherrer Institut, 5232 Villigen PSI, Switzerland}

\author{Mahesh Ramakrishnan}
\affiliation{Photon Science Division, Paul Scherrer Institut, 5232 Villigen PSI, Switzerland}

\author{Zoran Ristic}
\thanks{Current address: Department of Radiation Chemistry and Physics, VINCA Institute of Nuclear Sciences, Belgrade University, 11000 Belgrade, Serbia}
\affiliation{Photon Science Division, Paul Scherrer Institut, 5232 Villigen PSI, Switzerland}

\author{Valerio Scagnoli}
\affiliation{Laboratory for Multiscale Materials Experiments, Paul Scherrer Institut, 5232 Villigen PSI, Switzerland}
\affiliation{Laboratory for Mesoscopic Systems, Department of Materials, ETH Zurich, 8093 Zurich, Switzerland}

\author{Yoshikazu Tanaka}
\affiliation{RIKEN SPring-8 Center, Hyogo 679-5148, Japan}

\author{Tadashi Togashi}
\affiliation{RIKEN SPring-8 Center, Hyogo 679-5148, Japan}
\affiliation{Japan Synchrotron Radiation Research Institute, Hyogo 679-5198, Japan}

\author{Kensuke Tono}
\affiliation{RIKEN SPring-8 Center, Hyogo 679-5148, Japan}
\affiliation{Japan Synchrotron Radiation Research Institute, Hyogo 679-5198, Japan}

\author{Ivan Usov}
\affiliation{Science IT, Paul Scherrer Institut, 5232 Villigen PSI, Switzerland}

\author{Yoav W. Windsor}
\thanks{Current address: Fritz Haber Institute of the Max Planck Society, 14195 Berlin, Germany}
\affiliation{Photon Science Division, Paul Scherrer Institut, 5232 Villigen PSI, Switzerland}

\author{Makina Yabashi}
\affiliation{RIKEN SPring-8 Center, Hyogo 679-5148, Japan}
\affiliation{Japan Synchrotron Radiation Research Institute, Hyogo 679-5198, Japan}

\author{Steven L. Johnson}
\affiliation{Photon Science Division, Paul Scherrer Institut, 5232 Villigen PSI, Switzerland}
\affiliation{Institute for Quantum Electronics, ETH Z\"{u}rich, 8093 Z\"{u}rich, Switzerland}

\author{Paul Beaud}
\affiliation{Photon Science Division, Paul Scherrer Institut, 5232 Villigen PSI, Switzerland}

\author{Urs Staub}
\affiliation{Photon Science Division, Paul Scherrer Institut, 5232 Villigen PSI, Switzerland}

\begin{abstract}
The control of materials properties with light is a promising approach towards the realization of faster and smaller electronic devices. With phases that can be controlled \textit{via} strain, pressure, chemical composition or dimensionality, nickelates are good candidates for the development of a new generation of high performance and low consumption devices. Here we analyze the photoinduced dynamics in a single crystalline NdNiO$_3$ film upon excitation across the electronic gap. Using time-resolved reflectivity and resonant x-ray diffraction, we show that the pump pulse induces an insulator-to-metal transition, accompanied by the melting of the charge order. Finally we compare our results to similar studies in manganites and show that the same model can be used to describe the dynamics in nickelates, hinting towards a unified description of these photoinduced phase transitions.
\end{abstract}

\maketitle

\section{Introduction}
Transition metal oxides often display complex phase diagrams that originate from the tight interplay between structural, electronic and magnetic degrees of freedom. The boundaries of these phase diagrams can be tuned by external stimuli, allowing for an effective control of the properties of these functional materials. In particular, epitaxial growth of thin films are commonly used to alter the properties of the bulk material, which lead to creation of a 2 dimensional electron gas \cite{Ohtomo2004}, make a material polar \cite{Haeni2004} or even multiferroic \cite{Mundy2016}, or change its magnetic structure \cite{Windsor2014}. 
Simple perovskite nickelates \textit{R}NiO$_3$ (\textit{R}=Y or a Lanthanide ion) belong to a class of materials, which exhibit a insulator-to-metal transition (IMT) (except \textit{R}=La) and have an antiferromagnetic insulating ground state. They have a small or even negative charge transfer energy and are considered to belong to charge-transfer insulators \cite{Mizokawa1991}. The IMT has been found to be associated with the occurrence of charge order \cite{Alonso1999, Staub2002}, which leads to a monoclinic structural distortion and an alternation of the size of the  NiO$_6$ octahedra. No orbital order has been found at the Ni sites \cite{Scagnoli2005}, which is consistent with the non-collinear magnetic structure found by resonant magnetic x-ray scattering \cite{Scagnoli2006}. More recently, it has been theoretically predicted, that the charge order is localized in the Ni -- O bonds, leading to a splitting of the $d^8\underbar{\textit{L}}$ configuration, where $\underbar{\textit{L}}$ is an oxygen ligand hole, to a $d^8\underbar{\textit{L}}^2$ and $d^8$ state \cite{Mizokawa2000, Johnston2014} that is associated with the bond order of enlarged and compressed oxygen octahedra.

The IMT in perovskite nickleates is strongly affected by chemical composition, strain, and dimensional confinement or the proximity to other oxide layers \cite{Boris2011, Frano2013, Middey2016, Catalano2018}. Nickelates are also sensitive to impulsive photoexcitation and different approaches have been taken to control the electronic or magnetic properties of these materials with short light pulses. The temperature dependent response in NdNiO$_3$ was studied after photoexcitation of the electronic system \cite{Ruello2007, Ruello2009} and different responses were identified in the transient reflectivity signal above and below the ordering temperature. The melting of the antiferromagnetic order was also investigated with time-resolved x-ray diffraction \cite{Caviglia2013}. The reduction of magnetic order was shown to be linked to a rebalance of the charge on the NiO$_6$ sites and the dynamics of the Ni and Nd ions were found to be decoupled during this nonthermal process. The resonant excitation of a substrate vibrational mode with mid-infrared pulses also induces interesting dynamics \cite{Caviglia2012, Forst2015}. This ultrafast strain engineering was shown to trigger the melting of charge and antiferromagnetic order, as well as the relaxation of the structural distortion, each launched from the interface with different propagation velocities \cite{Forst2017}.

In this study, we use the pump-probe approach to explore the photoinduced insulator-to-metal transition in an epitaxially grown NdNiO$_3$ film after direct excitation of the electronic system. The sample is excited with short 800 nm pulses, corresponding to a transition across the gap in the antibonding $e_g$ band of nickel \cite{Stewart2011}. The transient dynamics are studied using two different approaches. The temperature and fluence dependence of the transient reflectivity is measured at a wavelength of 800 nm and a more direct view on the time evolution of the long range charge order is provided by time-resolved resonant x-ray diffraction.

\section{Experimental details}
\begin{figure}
\centering
\includegraphics[scale=1]{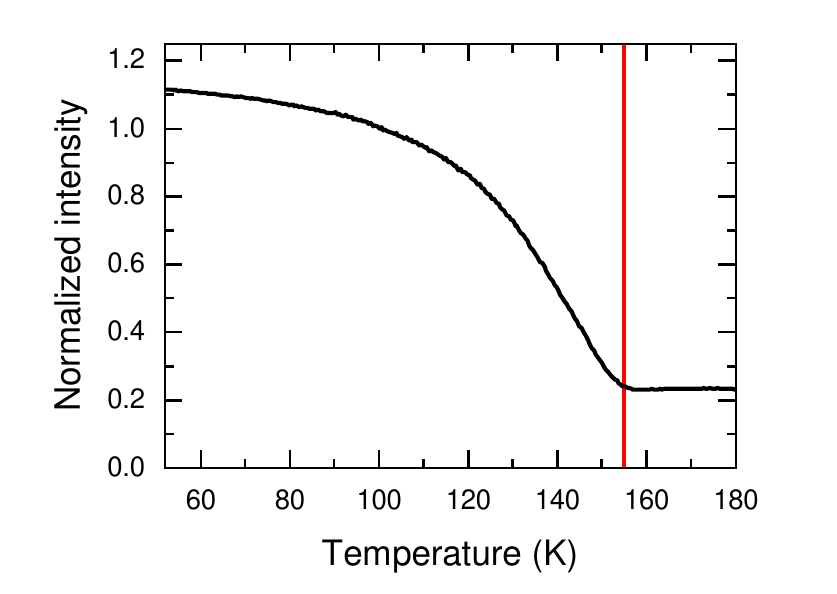}
\caption{Temperature dependence upon warming of the $(01\bar{3})$ reflection at resonance ($E=8.345$ keV). The diffracted intensity is normalized to the value at 100 K and the vertical line indicates the ordering temperature $T_{\text{CO}}=155$ K, about 50 K below that of the bulk material.}
\label{Tscan}
\end{figure}

The 60 nm thick NdNiO$_3$ thin film was grown by pulsed laser deposition on a $Pbnm$ $(110)$ NdGaO$_3$ substrate as described elsewhere \cite{Dhaka2015}. The sample was characterized at the X04SA surface diffraction beamline at the Swiss Light Source at PSI. The charge order can be directly measured by probing reflections of the type $(0kl)$ (in $Pbnm$ symmetry), with $k$ and $l$ odd, at the Ni $K$ edge resonance \cite{Staub2002, Scagnoli2005}. The ordering temperature measured \textit{via} the disappearance of the resonant contribution in the $(01\bar{3})$ is $T_\text{CO}=155$ K (Fig. \ref{Tscan}), about 50 K below the bulk value and consistent with other NdNiO$_3$ thin films \cite{Dhaka2015}.

The time-resolved x-ray diffraction experiments were carried out at the beamline 3 of the SACLA free electron laser (FEL) \cite{Ishikawa2012, Tono2013}. The x-ray energy was set to the Ni $K$ edge, granting sensitivity to the charge ordering pattern in the low temperature phase. The x-ray beam was focused with a KB mirror system to $50 \times 40$ $\mu$m$^2$ (full width at half maximum) and the 800 nm pump laser was focused to $250 \times 250$ $\mu$m$^2$. Taking into account the incident angles ($10^{\circ}$ for the the pump laser and $5^{\circ}$ for the x-ray probe) of the beams, the footprint of the x-ray beam was about 2 times smaller than the laser beam, ensuring homogeneous in-plane excitation. The diffracted signal was monochromatized with a Cu (111) analyzer by choosing the $(222)$ reflection that results in a $2\theta\approx 90^{\circ}$ at the nickel $K$ edge, and measured with a YAP detector. The energy was optimized to the maximum of the $(01\bar{3})$ reflection resonance that has a very similar energy dependence as the $(015)$ at 8.345 keV \cite{Staub2002}. Despite being at resonance, the background remains actually very small because the fluorescence has a different energy and is therefore suppressed by the analyzer crystal. Because the monochromator (analyzer) is placed downstream from the I$_0$ monitor, we are, however, very sensitive to the shot-to-shot intensity jitter of that particular spectral component. The sample temperature was kept at about 100 K, well below the charge ordering temperature, by means of a nitrogen cryoblower. Due to the jitter between the pump and probe pulses, the overall time-resolution was about 700 fs, significantly lower than the cross correlation between optical pump and x-ray probe of 50 fs. Note that a timing diagnostic tool, allowing to correct for the jitter was installed at the beamline in the meantime \cite{Katayama2016}.

Time-resolved reflectivity was performed with a 800 nm probe at a repetition rate of 2 kHz, alternating between pumped and unpumped shots, in order to correct for slow drifts in the laser and electronics. The pump and probe beams were produced by splitting the output of a regenerative amplifier system seeded by a Ti:S oscillator. The pump was focused to $480 \times 550$ $\mu$m$^2$ and the probe to $110 \times 110$ $\mu$m$^2$, guaranteeing homogeneous excitation. The time resolution, measured from the cross-correlation of the two beams in a Ba(BO$_2$)$_2$ crystal, was about 80 fs. The sample was placed in a cryostat coupled to a close-loop helium compressor, allowing temperature stabilization down to approximately 5 K. The reflectivity was measured at normal incidence with a fast photodiode and gated by a boxcar integrator.

\section{Results}

\subsection{Transient optical reflectivity}
\subsubsection{Temperature dependence}
\begin{figure*}
\centering
\includegraphics[scale=1]{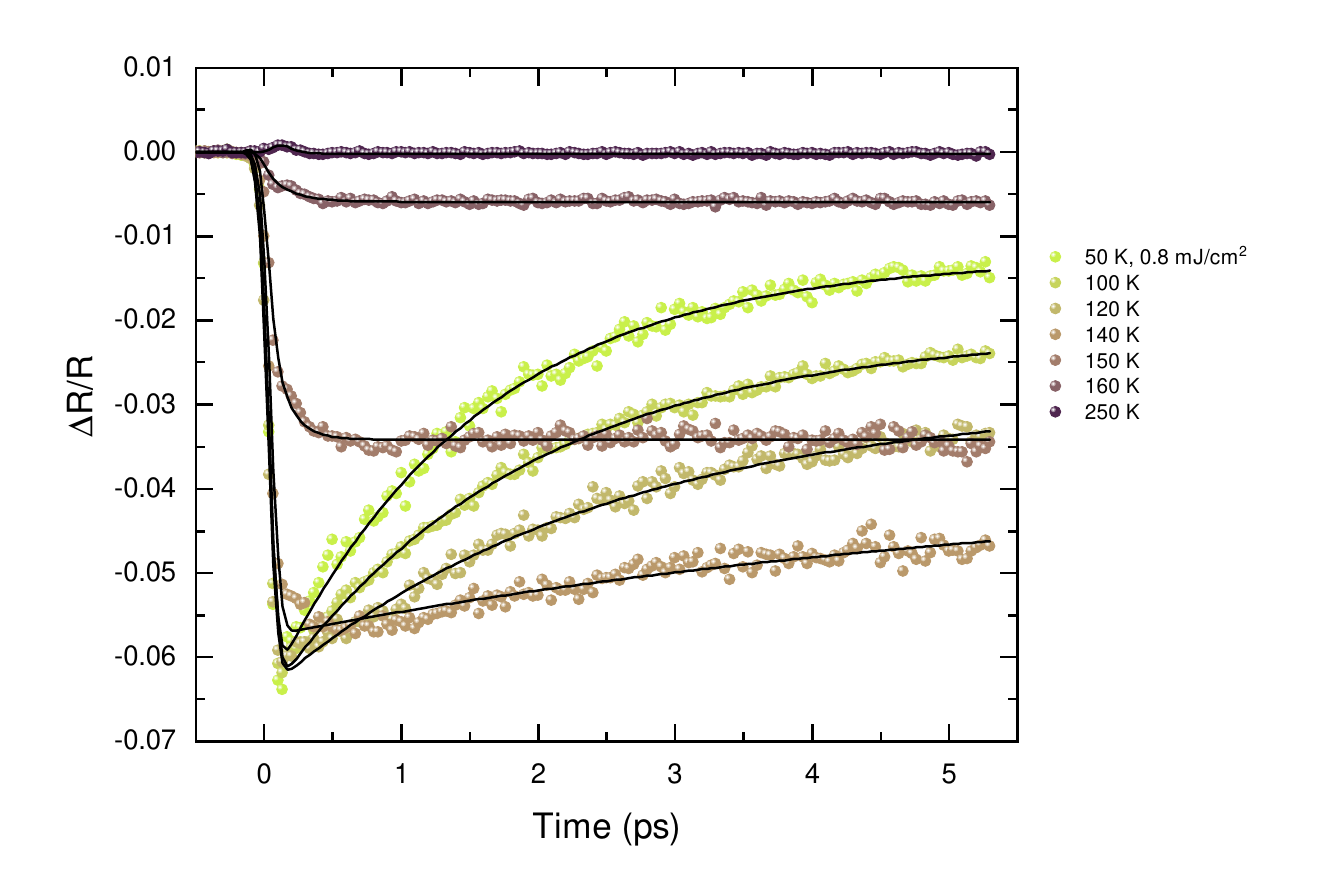}
\caption{Temperature dependence of the transient reflectivity in NdNiO$_3$. Major changes are observed around the critical temperature $T_{\text{CO}}=150$ K. The black lines are fit with Eq. \ref{opt_fit}.}
\label{optical_tdep}
\end{figure*}

The temperature dependence of the 800 nm transient reflectivity changes at a fluence of 0.8 mJcm$^{-2}$ are shown in Fig. \ref{optical_tdep}. For the sake of clarity, only a few selected temperatures are displayed out of a larger data set. Very different responses are observed when exciting the sample above or below $T_{\text{CO}}$. Below the ordering temperature, the reflectivity decreases promptly upon excitation followed by a slower relaxation towards a metastable state. The reduced reflectivity originates from a transfer of spectral weight to lower frequencies linked to the increased DC conductivity after photo-doping. Above $T_{\text{CO}}$, a slight and prompt increase in reflectivity is observed and decays within a few hundreds of femtoseconds back to the equilibrium value. This initial increase of reflectivity is consistent with the response of a metallic state \cite{Schoenlein1987}, but its rapid recovery resembles the dynamics observed in semi-conductors \cite{Gupta1992}. Finally, an intermediate regime is observed at the transition temperature with a prompt decrease followed by a slower decrease after which either reflectivity is constant or recovers, dependent on the starting temperature.

\begin{figure*}
\centering
\includegraphics[scale=1]{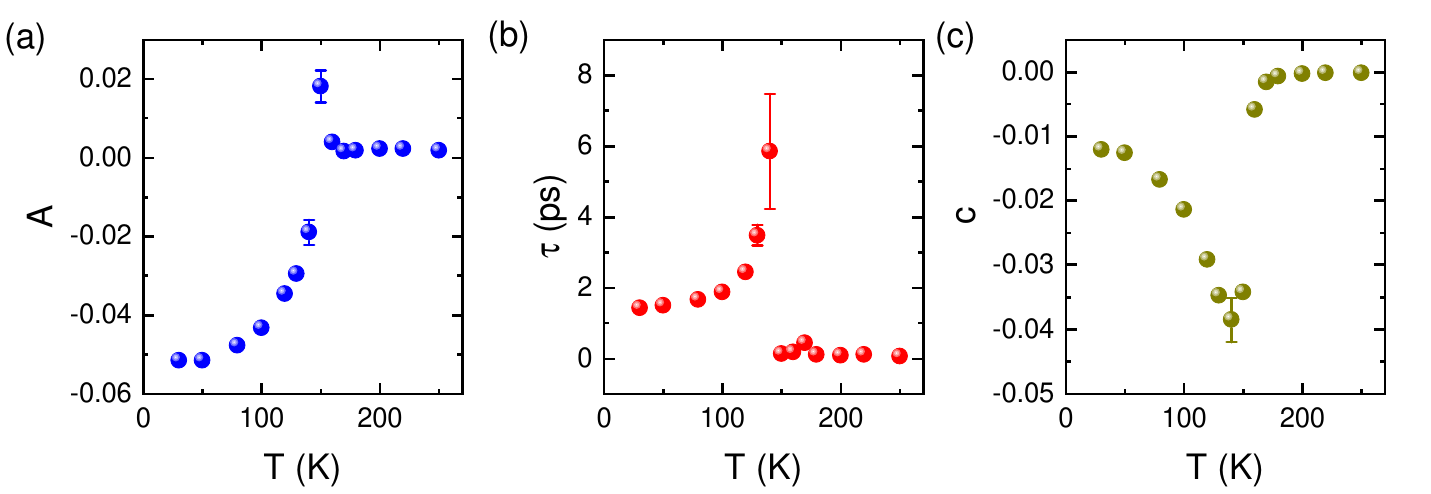}
\caption{Parameters from the fit of Eq. \ref{opt_fit} to the data in Fig. \ref{optical_tdep} as a function of temperature. The transition around 150 K is clearly visible in all three parameters.}
\label{optical_tdep_param}
\end{figure*}

This behaviour is best seen when fitting the data with a single exponential relaxation
\begin{equation}
\frac{\Delta R}{R} = 0.5 \cdot \left( \textit{erf}\left( \frac{t-t_0}{\sigma}\right) +1 \right) \cdot \left( Ae^{-\frac{t-t_0}{\tau}} + c \right)\text{.}
\label{opt_fit}
\end{equation}
The first part simulates the time-dependent excitation profile with duration of $\sigma=80$ fs and arrival time $t_0$, with $\textit{erf}(x)$ being the error function. The initial reflectivity change is given by the amplitude $A$, the timescale of the recovery by $\tau$ and the constant $c$ accounts for the long-lived transient that last well beyond 5 ps. The fitted curves are overlaid to the data in Fig. \ref{optical_tdep} (black lines). The fitted parameters are reported in Fig. \ref{optical_tdep_param}. They all display a sharp discontinuity in their respective trend around the critical temperature $T_{\text{CO}}$. In particular, the amplitude $A$ changes sign at the transition. The recovery slows down as the temperature is increased in the insulating phase and becomes suddenly extremely fast in the metallic state. Close to the critical temperature, the fast drop in reflectivity is followed by a further reduction in reflectivity in a time period of about 0.5 ps and the recovery is suppressed. These additional features leads to the increased error bars in Fig. \ref{optical_tdep_param}, visualizing the difficulties to fit the data with this simple model. These effects may be caused by the thermalization between co-existing metallic and insulating domains in the hysteretic window \cite{Post2018}. Additionally, the pump penetration depth is of the order of the film thickness and the top part of the film is thus more excited than deeper layers. Close to the critical temperature, this increased excitation may trigger the transition to the high temperature phase and the further reduction of reflectivity can also be attributed to thermalization between melted and unmelted layers.

It is worth noting that no coherent oscillations are observed in the low temperature phase, as opposed to the charge ordered manganites \cite{Lim2005, Matsuzaki2009, Caviezel2012}. A structural distortion occurs nonetheless at the ordering temperature, leading to a lowering of the crystal symmetry from orthorhombic to monoclinic. In the ultrafast regime, structural dynamics are decoupled between the small atomic motions within the unit cell and the change of the unit cell shape and/or size \cite{Esposito2018}. The first part is generally faster and carried by coherent optical phonons, while the second part is limited by the speed of sound in the material. In nickelates, the main structural motion involves the expansion/contraction of the oxygen octahedra. This motion is driven by breathing modes, whose frequency cannot be resolved with the experimental time resolution of 80 fs, but should eventually lead to the displacement of the heavier rare-earth ions, at lower frequencies. The amplitude of the rare-earth displacements during the thermodynamic transition are comparable to those in manganites \cite{Garcia-Munoz2009}, where a pronounced coherent oscillation is observed upon photoexcitation. The absence of coherent oscillation at low frequency in the refelctivity is thus puzzling. Several hypothesis can, however, explain our observations. The phonon may be overdamped, or the motion of the rare-earth ions not be strongly coupled to the oxygen octahedra breathing mode and only follows the slower change of the unit cell involving the monoclinic to orthorhombic transition, which has been reported upon mid-infrared excitation \cite{Forst2017}. Finally, the rare-earth motion may only induce small variation of the refractive index. The transient reflectivity signal would then be very little sensitive to those motions.

A similar pump-probe study at 760 nm has been reported on a 150 nm NdNiO$_3$ film grown on a $(100)$ silicon substrate, corresponding to the $(110)$ orthorhombic direction \cite{Ruello2007}. In this study, the reflectivity was found to increase after excitation in both the insulating and the metallic phases. Moreover, a crossover from a double to a single exponential recovery is observed at the transition. Our data are thus consistent with the reported behavior in the metallic state, but differ significantly in the ordered phase. This might be caused by the microstructure difference of films grown on silicon substrate, which have much larger strain compared to films grown on orthorhombic perovskite oxides\cite{Laffez2000}. Nickelates are indeed extremely sensitive to strain \cite{Catalano2018} and, for large strains, it is expected that the films break up in smaller domains\cite{Laffez2000}. The increase number of domain walls, which possibly remain conductive \cite{Post2018}, could lead to a different overall response in the insulating state.

\subsubsection{Fluence dependence}
\begin{figure*}
\centering
\includegraphics[scale=1]{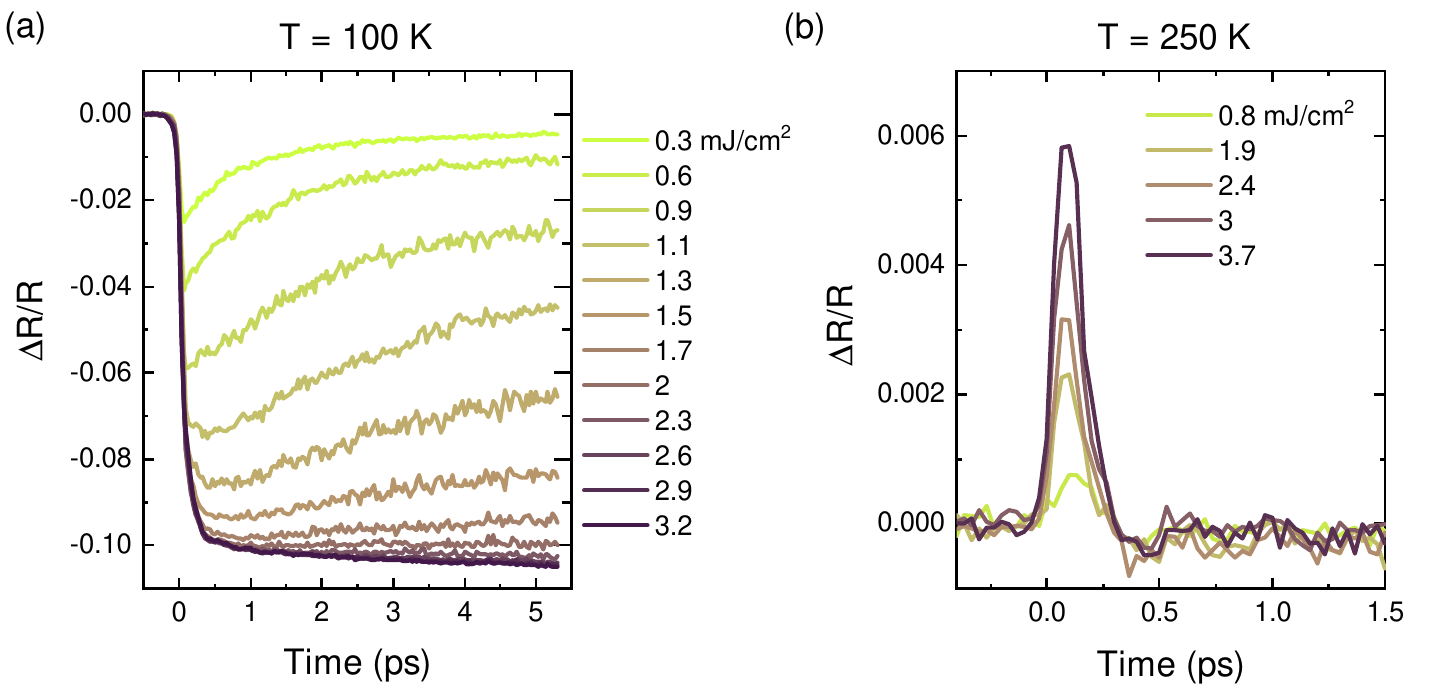}
\caption{Transient 800 nm reflectivity as a function of fluence for temperature below (100 K) (a) and above (250 K) (b) the ordering temperature.}
\label{optical_fludep}
\end{figure*}

The fluence dependent transient reflectivity above and below the ordering temperature is shown in Fig. \ref{optical_fludep}. Panel (a) shows the transient reflectivity at 100 K, well below the ordering temperature. At low fluence, the prompt intensity drop is followed by a slower relaxation. For fluences above 2 mJcm$^{-2}$, the initial drop saturates to about $-10\%$. At this point, no more recovery is observed within the first 5 picoseconds and the reflectivity seems actually to drop even further on longer timescales. Based on the optical properties at 800 nm (1.55 eV) for NdNiO$_3$ grown on NdGaO$_3$ $(110)$, a $10\%$ reduction in reflectivity is expected when heating through the transition \cite{Ruppen2015}. The saturation of the transient reflectivity drop at this value clearly indicates the completion of a photoinduced IMT in this material.

For fluences above 1.1 mJcm$^{-2}$, the reflectivity drops further on a 0.5 ps time scale. These dynamics resemble those close to the critical temperature (Fig. \ref{optical_tdep}), possibly indicating the presence of phase co-existence. Because of the inhomogeneities in the film, it is, indeed, possible that parts of the film undergo the transition at a lower fluence.

Close to the critical temperature, the fluence needed to induce the transition seems to decrease. Indeed, the data at 140 K and 150 K of the transient reflectivity (Fig. \ref{optical_tdep}) resemble the curves around the critical fluence at 100 K. The fluence of 0.8 mJcm$^{-2}$ appears to be large enough to induce the transition at these temperatures. At high temperature, the initial increase of reflectivity scales linearly with fluence (Fig. \ref{optical_fludep} (b)) and the recovery time of a couple of 100 fs does not show any significant fluence dependence.

The dynamics above the critical fluence resembles closely those observed upon strong resonant excitation of a substrate lattice mode \cite{Caviglia2012}. This type of excitation was shown to also induce the melting of the charge order and the relaxation of the structural distortion, demonstrating a transition to a metastable state resembling the high temperature phase \cite{Forst2015, Forst2017}. The fact that the same response is observed here indicates that the same transient state is being induced.

\subsection{Resonant x-ray diffraction}
\begin{figure}
\centering
\includegraphics[scale=1]{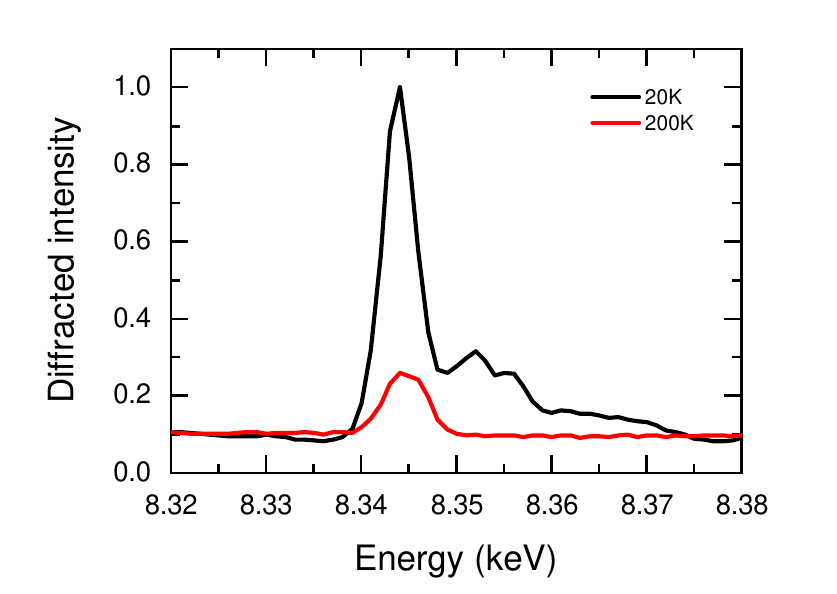}
\caption{Energy-dependent XRD intensity of the $(01\bar{3})$ forbidden reflection above and below $T_{\text{CO}}$. The intensity is normalized to the maximal intensity of the low temperature scan.}
\label{escan}
\end{figure}

Resonant x-ray diffraction provides a more quantitative and direct view of this IMT, as it can directly probe the underlying electronic ordering phenomena. In Fig. \ref{escan} an energy scan of the $(01\bar{3})$ forbidden reflection above and below $T_{\text{CO}}$ is shown. The reflections of the type $(0kl)$ with $k$ and $l$ odd are forbidden in the high temperature orthorhombic symmetry and are directly sensitive to the charge order at resonance in low temperature phase. For $T>T_\text{CO}$ and far from the nickel $K$ edge (8.33 keV), the peak intensity is almost zero and the small energy independent contribution may originate from a slight distortion of the orthorhombic structure, possibly stabilized by strain. There is also a small contribution from the substrate, possibly multiple scattering, as verified by measuring a bare NdGaO$_3$ crystal. At resonance, there is a small orbital contribution left as studied in detail in Ref. \cite{Scagnoli2005}. Below $T_\text{CO}$, the resonant contribution is significantly enhanced due to the charge ordering at the NiO$_6$ octahedra \cite{Staub2002, Scagnoli2005}. The high temperature contribution to the reflection intensity is about 25\% to the one at 100 K.

\begin{figure*}
\centering
\includegraphics[scale=1]{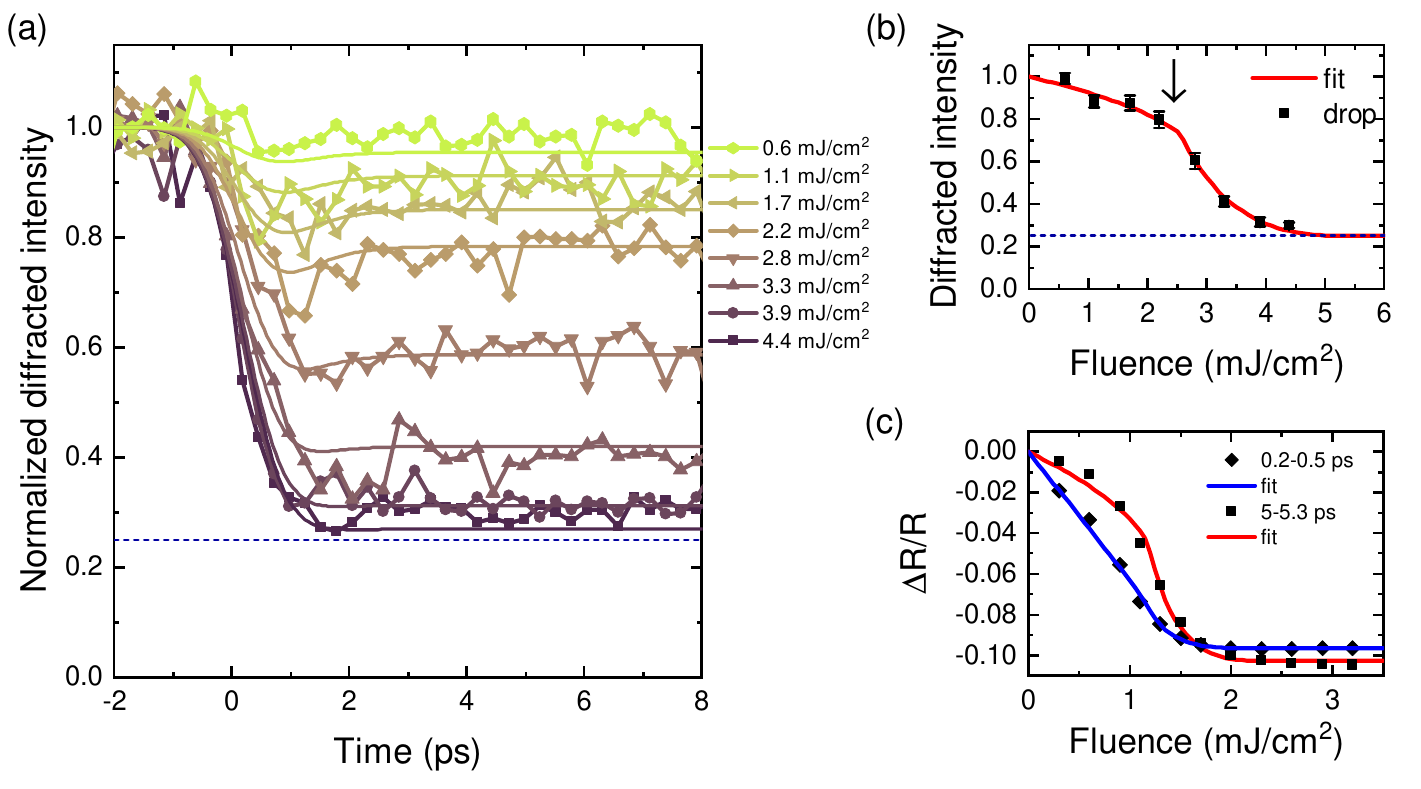}
\caption{
(a) Time dependence of the $(01\bar3)$ reflection after excitation with 800 nm pulses. The horizontal dashed line indicate the intensity of the peak above the ordering temperature.
(b) Average normalized intensity between 4 and 6 ps as a function of fluence. The black arrow indicates the fluence corresponding to the critical energy density at the surface ($f_c=2.5$ mJcm$^{-2}$).
(c) Average reflectivity drop at early and later times after excitation. The data are fitted with Eq. \ref{drop_fit}.
}
\label{xray_fludep}
\end{figure*}

The dynamics of the $(01\bar3)$ reflection as a function of time is shown in Fig. \ref{xray_fludep} (a). As stated in the experimental section, this reflection at resonance is a direct measure of the long-rang order of the charge order. A fast reduction of intensity is observed immediately after photo-excitation with the pump pulse, evidencing a prompt reduction of the charge order. Above 3.3 mJ/cm$^2$ the intensity drop saturates to about 30\% of its maximum intensity. This corresponds closely to the remaining diffraction intensity observed in the high temperature phase (Fig. \ref{escan}), demonstrating the complete melting of the charge ordered state.

In Fig. \ref{xray_fludep} (b), the intensity drop averaged between 4 and 6 ps is plotted as a function of the incoming fluence. In the equilibrium phase, the enhanced resonant contribution of this forbidden peak is an order parameter of the charge ordered phase. Assuming that the electronic system has thermalized, the transient intensity of the $(01\bar3)$ can be considered as a valid order parameter square of the electronic order. The intensity of the reflection as a function of fluence is thus described with a continuous Landau-like order parameter, a model based on the description of the charge order dynamics in a manganite \cite{Beaud2014}. As the excitation depth ($\approx 50$ nm) compares to the film thickness and is much smaller than the x-ray penetration length, we account for the depth-dependent excitation profile by splitting the sample in $N$ layers. Each layer contributes to the diffracted intensity according to its excitation density $n_i=n_0 e^{-z/z_0}$, where $z_0$ is the penetration depth of the laser. The diffracted intensity is then given by
\begin{equation}
I=\eta^2, \quad \text{with }
\eta = \frac{1}{N} \sum_i^N
\begin{cases}
	\left( 1- \frac{n_i}{n_c} \right)^{\gamma}, & \text{if } n_i<n_c \\
	0, & \text{otherwise.}
\end{cases}
\label{drop_fit}
\end{equation}
An additional scaling factor and an offset are used to account for the remaining intensity at high temperature and are taken from the intensity above and below the ordering temperature in Fig. \ref{escan}. The fit yields a critical energy density $n_c = 343 \pm 7$ Jcm$^{-3}$ and an exponent $\gamma = 0.15 \pm 0.02$. In Figure \ref{xray_fludep} (c), the reflectivity drop extracted from Fig. \ref{optical_fludep} (a) is reported for early and later delays after excitation. Just after excitation, the reflectivity drop is linear. At later delays, its response qualitatively resembles that of the charge order. Fit of these curves with Eq. \ref{drop_fit} yields a critical energy density $n_c = 165 \pm 5$ Jcm$^{-3}$ and $\gamma = 0.18 \pm 0.02$. The difference between the critical energy densities found for the two experiments is intriguing. The critical energy density in the optical experiment corresponds to a critical fluence at the surface of 1.2 mJcm$^{-2}$, which is compatible with that found for the melting of the magnetic order \cite{Caviglia2013}. The critical energy density for the melting of the charge order is with approximately twice as large. However, the uncertainties in the laser power and beam size measurements in these experiments as well as the different pump pulse might be responsible for these differences.

In manganites, the photoinduced melting of the charge order and its associated dynamics were successfully described with a time-dependent order parameter \cite{Beaud2014}. Starting from Eq. \ref{drop_fit}, we account for the electron-lattice thermalization by considering the time-evolution of the energy deposited in the electronic system: $n_i \rightarrow n_i(t)$. Indeed after excitation, the excess energy dissipates into the lattice until a common temperature is reached. Below the critical energy density $n_c$, this dissipation leads to a partial recovery of the charge order and its order parameter. Above $n_c$, the transition occurrs and the system remains trapped in an electronically disordered metastable state. It was proposed that this description could apply to other photoinduced transitions as well. We have thus applied this model to the case of nickelates. We have considered the empirical evolving energy density $n_i(t)$ that was also used for the manganites \cite{Beaud2014}:
\begin{equation}
n_i(t) = (n_{i0}-\alpha n_c) e^{-t/\tau} + \alpha n_c
\end{equation}
with 
\begin{equation}
\alpha = 1- \left(1-\frac{n_{i0}}{n_c}\right)^{2 \gamma}
\end{equation}
where $n_{i0}$ is the energy deposited by the laser at $t=0$ in layer $i$, $\tau$ is the timescale of the thermalization and $\gamma$ is a parameter that determines the amount of order in the metastable state. Only three free parameters are used to fit all curves simultaneously ($n_c$, $\tau$ and $\gamma$), and the resulting fits are overlaid to the data in Fig. \ref{xray_fludep} (a). From the reasonable good agreement between the model and the data, we conclude that the time-dependent order parameter model may also be used to describe the melting of charge order in nickelates, hinting at a possible universality of this approach.

\section{Conclusion}
In summary, we have shown that the temperature-dependent phase transition is clearly identifiable in the reflectivity changes. The transient response is consistent with an insulating phase at low temperature and with a metallic state above the ordering temperature.
The fluence-dependent transient reflectivity reveals the completion of a photoinduced insulator-to-metal transition above a certain threshold.
A detailed investigation of the charge order dynamics with resonant x-ray diffraction shows indeed a melting of the charge order above a critical fluence, confirming the photoinduced phase transition. The description of the data with a time-dependent order parameter model demonstrated that, as for manganites, these dynamics can be well described using exclusively the absorbed energy.

\section{Acknowledgments}
This work was supported by the Swiss National Science Foundation and its National Centers of Competence in Research in Molecular Ultrafast Science and Technology (NCCR MUST). E.M.B. acknowledges funding from the European Community’s Seventh Framework Programme (FP7/2007-2013) under Grant Agreement No. 290605 (PSI-FELLOW/COFUND).
The time-resolved x-ray experiments were performed under the approval of the Japan Synchrotron Radiation Research Institute (JASRI proposals No. 2014A8006 and 2015B8009).

\bibliography{./refs_NNO}

\end{document}